# Nonvolatile Multi-level Memory and Boolean Logic Gates Based on a Single Memtranstor

Jianxin Shen, Dashan Shang[†], Yisheng Chai, Yue Wang, Junzhuang Cong, Shipeng Shen, Liqin Yan, Wenhong Wang, and Young Sun[*]

Beijing National Laboratory for Condensed Matter Physics, Institute of Physics, Chinese Academy of Sciences, Beijing 100190, P. R. China
[*]youngsun@iphy.ac.cn; [†]shangdashan@iphy.ac.cn

Abstract

Memtranstor that correlates charge and magnetic flux via nonlinear magnetoelectric effects has a great potential in developing next-generation nonvolatile devices. In addition to multi-level nonvolatile memory, we demonstrate here that nonvolatile logic gates such as NOR and NAND can be implemented in a single memtranstor made of the Ni/PMN-PT/Ni heterostructure. After applying two sequent voltage pulses ($X_1$, $X_2$) as the logic inputs on the memtranstor, the output magnetoelectric voltage can be positive high (logic "1"), positive low (logic "0"), or negative (logic "0"), depending on the levels of $X_1$ and $X_2$. The underlying physical mechanism is related to the complete or partial reversal of ferroelectric polarization controlled by inputting selective voltage pulses, which determines the magnitude and sign of the magnetoelectric voltage coefficient. The combined functions of both memory and logic could enable the memtranstor as a promising candidate for future computing systems beyond von Neumann architecture.

## I. INTRODUCTION

In electric circuit theory, the three well-known fundamental elements (resistor, capacitor, and inductor) are defined from the linear relationship between two of the four basic circuit variables (charge $q$, current $i$, voltage $v$, and magnetic flux $\varphi$), as schematically illustrated in Fig. 1(a). The fourth fundamental linear element that is defined directly from the $q$-$\varphi$ relationship has been under debate for many years [1-4]. Recently, we demonstrated that the true fourth element can be realized by employing the magnetoelectric (ME) effects [4,5], *i.e.*, magnetic field control of electric polarization ($P$) and electric field control of magnetization ($M$) [6-9], because a direct relationship between $q$ and $\varphi$ is built up in a simple two-terminal passive device via the ME coupling, as illustrated in Fig. 1(b). The fourth fundamental element based on the ME effects is termed transtor. Corresponding to the linear elements, there are four nonlinear memelements [10], namely, memristor, memcapacitor, meminductor, and memtranstor. These memelements provide a great potential to broaden electric circuit functionality for next-generation electronic devices. For instance, the memristor is attracting tremendous interests because of its substantial technological applications [11-18]. Similarly, the memtranstor also has a great promise to develop advanced electronic devices.

The memtranstor is characterized by nonlinear hysteresis loops shown in Fig. 1(c) and 1(d). Either the butterfly-shaped or the pinched hysteresis loops, depending on the magnitude of external stimulus, provide the basis of a new concept of nonvolatile memory. The principle is to utilize the different states of transtance, $T=dq/d\varphi$, or equivalently, the ME voltage coefficient $\alpha_E=dE/dH$, varying from positive to negative (Fig. 1(c)) or high to low (Fig. 1(d)), to store binary information. In our recent works [19,20], both two-level and multi-level nonvolatile memory devices have been demonstrated based on the PMN-PT/Terfenol-*D* memtranstor. In this work, we show that the functionality of the memtranstor can be further exploited to implement nonvolatile logic gates.

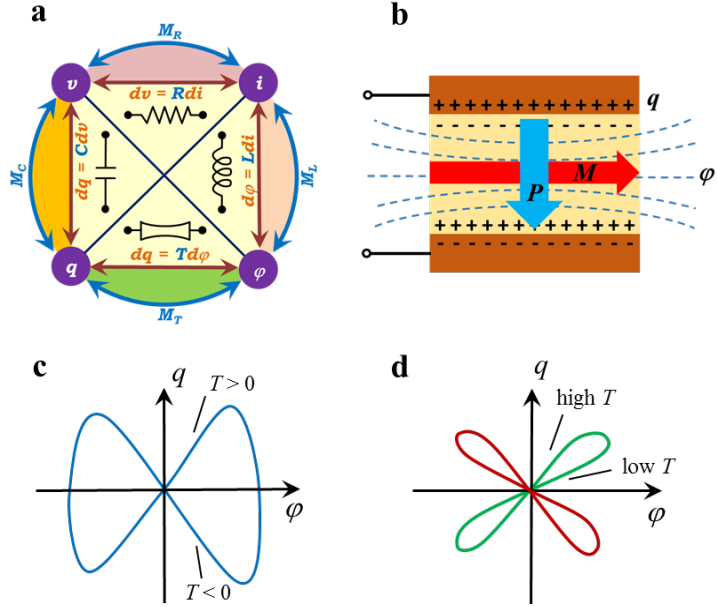

**Figure 1.** The Principle of the memtranstor. (a) The full diagram of fundamental circuit elements. The four linear elements (resistor, capacitor, inductor, and transtor) are defined by the linear relationships between two of four basic variables ($v$, $i$, $q$, $\varphi$), respectively. The four memelements (memristor, memcapacitor, meminductor, and memtranstor) are defined by the nonlinear relationships between two variables, respectively. (b) The typical structure of a memtranstor. The $q$-$\varphi$ relationship is realized via the ME coupling between magnetization ($M$) and polarization ($P$). (c) & (d) The characteristics of memtranstors. Either butterfly-shaped or pinched hysteresis loops can be observed, depending on the magnitude of external stimulus. The different states of transtance, $T=dq/d\varphi$, or equivalently, the ME voltage coefficient $\alpha_E=dE/dH$, ranging from positive to negative and high to low, are used to implement both nonvolatile memory and logic functions.

## II. EXPERIMENTS

The memtranstor used in this study is made of the Ni/PMN-PT/Ni heterostructure with in-plane $M$ and out-of-plane $P$, as illustrated in Fig. 2(a). The top and bottom Ni layers act as not only the magnetic components of the memtranstor but also the electrodes. The memtranstor is prepared by magneton sputtering Ni layers on the surfaces of $0.7Pb(Mg_{1/3}Nb_{2/3})O_3$-$0.3PbTiO_3$ (PMN-PT) single-crystal substrate with (110)-cut. Each Ni layer is 1 μm in thickness and the PMN-PT single crystal is 200 μm in thickness.

A conventional dynamic technique is employed to measure the ME voltage coefficient ($\alpha_E$). A Keithley 6221 AC source is used to supply an ac current to a solenoid to generate a small ac magnetic field $h_{ac}$ at a frequency of 10 kHz. In response, a synchronized 10 kHz ac ME voltage, $V_{ac}=x+y\mathrm{i}$, across the electrodes was measured by a lock-in amplifier (Stanford Research SR830). The ME voltage coefficient $\alpha_E$ is calculated by $\alpha_E = x/(h_{ac}t)$, where $t$ is the thickness of the ferroelectric layer (200 μm). To switch or pre-pole the electric polarization of PMN-PT, a Keithley 6517B electrometer was used to apply voltage pulse across the electrodes. The device was loaded in an Oxford TeslatronPT superconducting magnet system to apply the dc bias magnetic field ($H_{dc}$). All the measurements were performed at room temperature.

## III. RESULTS AND DISCUSSION

Figure 2(b) shows the ME voltage coefficient $\alpha_E$ of the Ni/PMN-PT/Ni memtranstor as a function of in-plane dc magnetic field. Before measuring $\alpha_E$, the memtranstor was pre-poled to set the direction of $P$ pointing upward or downward. The magnitude and sign of $\alpha_E$ depend on the relative orientation between $M$ and $P$: when the direction of $P$ is fixed, $\alpha_E$ can be controlled by reversing $M$ with in-plane magnetic field; when the direction of $M$ is fixed, the states of $\alpha_E$ can be controlled by fully or partially reversing $P$ with vertical electric fields. The latter case is employed to implement both nonvolatile memory and logic functions.

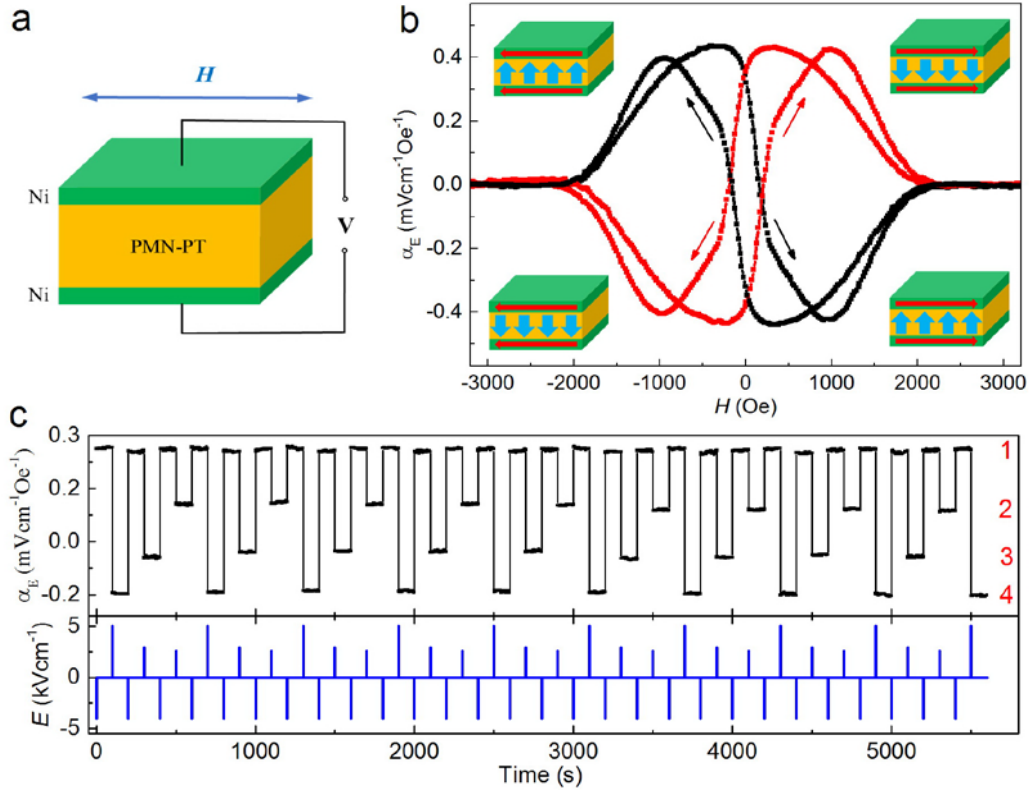

**Figure 2.** Multi-level nonvolatile memory based on the Ni/PMN-PT/Ni memtranstor. (a) The structure of the device and the measurement configuration. Two layers of Ni were deposited on the surfaces of PMN-PT(110) single crystal. The electric field was applied vertically to switch the direction of $P$ and the magnetic field was applied in plane to alter the direction of $M$. (b) The ME voltage coefficient $\alpha_E$ as a function of dc magnetic field with the PMN-PT layer pre-poled to $+P_s$ and $-P_s$, respectively. The states of $\alpha_E$ depend on the relative orientation between $M$ and $P$. The large hysteresis with a high remnant $\alpha_E$ at zero dc bias is beneficial for practical applications. (c) Repeatable multilevel switch of $\alpha_E$ by applying selective voltage pulses ($E=$ $-4.0$, $5.0$, $2.9$, and $2.6$ kVcm$^{-1}$), in zero dc bias magnetic field. After each voltage pulse (10 ms) $\alpha_E$ was measured for 100 s.

Compared with that of the PMN-PT/Terfenol-$D$ memtranstor [19], $\alpha_E$ of the Ni/PMN-PT/Ni memtranstor exhibits a broader hysteresis loop and a higher remanence at zero bias field. These features are favorable for practical applications because no dc bias is required. As seen in Fig. 2(c), the states of $\alpha_E$ can be well controlled by applying selective electric-field pulses ($E=-4.0$, $5.0$, $2.9$, and $2.6$ kVcm$^{-1}$) to fully or partially reverse $P$. Consequently, four clearly-separated states of $\alpha_E$ ranging from positive to negative are obtained repeatedly for many cycles, demonstrating a multi-level nonvolatile memory based on the Ni/PMN-PT/Ni memtranstor. As we discussed in our previous papers [19,20], the nonvolatile memory based on the memtranstors retains all the advantages of ferroelectric memory (simple structure, fast speed, low power consumption), but overcomes the drawback of destructive reading of *P*.

Furthermore, the memtranstor also provides the opportunity to build up novel computing systems beyond classical von Neumann architecture. In the past decade, many efforts have been devoted to develop future computing systems with logic-in-memory architecture where memory and logic gates are tightly integrated. Various nonvolatile devices including the memristors, magnetic tunneling junctions (MTJs), phase change memories, *etc.*, have been employed to implement nonvolatile logic functions [21-27]. In the following, we propose and demonstrate that the memtranstor also has the potential to implement nonvolatile logic gates.

Figure 3(a)-3(c) illustrate how to implement NOR logic function using a single memtranstor. NOR is a universal Boolean logic function of two binary inputs, where the output is always "0", except for the inputs being both "0", in which case the output is "1". The logic operation is accomplished by three stages: initialization, computation, and readout. First, the initializing operation prepares the memtranstor in the positive high $\alpha_E$ state. Then, the computing stage consists of inputting a sequence of two voltage pulses $X_1$ and $X_2$, carrying the input logic signals. The input voltage pulses with high and low magnitude are defined as logic "1" and "0", respectively. Last, the computation is read out by applying a small magnetic field to generate the output voltage via the ME effect, $\alpha_E = \frac{dE}{dH} = \frac{1}{t}\frac{dV}{dH}$, where *t* is the thickness of the ferroelectric layer.

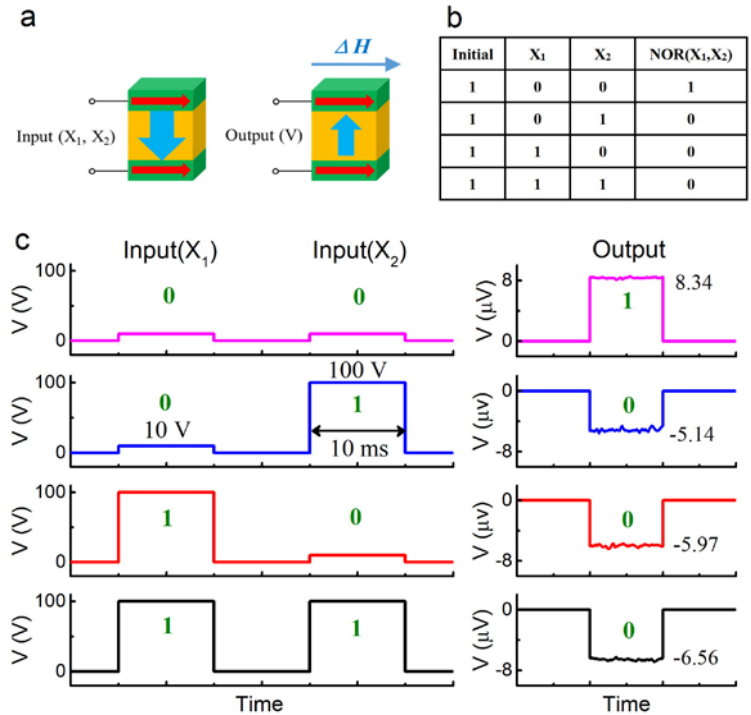

**Figure 3.** Nonvolatile NOR logic based on a single memtranstor – the first mode. (a) The schematic of the device structure and operations. After initialization, two sequent voltage pulses $X_1$ and $X_2$ were inputted into the device to do the computation. The low input (10 V) is set as logic "0" and the high input (100 V) as "1". The result is read out by applying a small in-plane magnetic field ($\Delta H$) to generate the output voltage via the ME effect. (b) The truth table of NOR operation. (c) Experimental results obtained on the Ni/PMN-PT/Ni memtranstor demonstrating the NOR operation. The output positive ME voltage is set as logic "1" and negative as logic "0". Inputting "0" does not alter the direction of *P* so that the output ME voltage remains

positive high. Inputting "1" fully reverses the direction of P so that the output ME voltage becomes negative.

As shown in Fig. 3(c), the experimental results obtained on the Ni/PMN-PT/Ni memtranstor confirm the NOR logic function. For $X_1$ and $X_2$ being both low, no change occurs in the memtranstor because the low inputs (10 V) do not alter the direction of P. Thus, $\alpha_E$ remains positive high, *i.e.*, the output ME voltage is positive high. When either $X_1$ or $X_2$ are high, the majority of P is reversed by the high input (100 V) so that $\alpha_E$ becomes negative. When both $X_1$ and $X_2$ are high, the direction of P is nearly fully reversed and $\alpha_E$ becomes negative high. In this case, we can define the positive and negative output ME voltage as logic "1" and "0", respectively, and the computations fulfill the truth table of the NOR logic.

Alternatively, the NOR logic can be implemented by another mode with lower input voltages. As shown in Fig. 4, after initialization, when both $X_1$ and $X_2$ are low (10 V), the output ME voltage remains positive high because no change occurs in the memtranstor. When one of $X_1$ and $X_2$ is high (60 V), a large proportion of P is reversed so that $\alpha_E$ is reduced from positive high to positive low. When both $X_1$ and $X_2$ are high (60 V), the majority of P is reversed and $\alpha_E$ becomes negative. In this scenario, the high output ME voltage is defined as logic "1", and the low or negative output ME voltage is defined as logic "0".

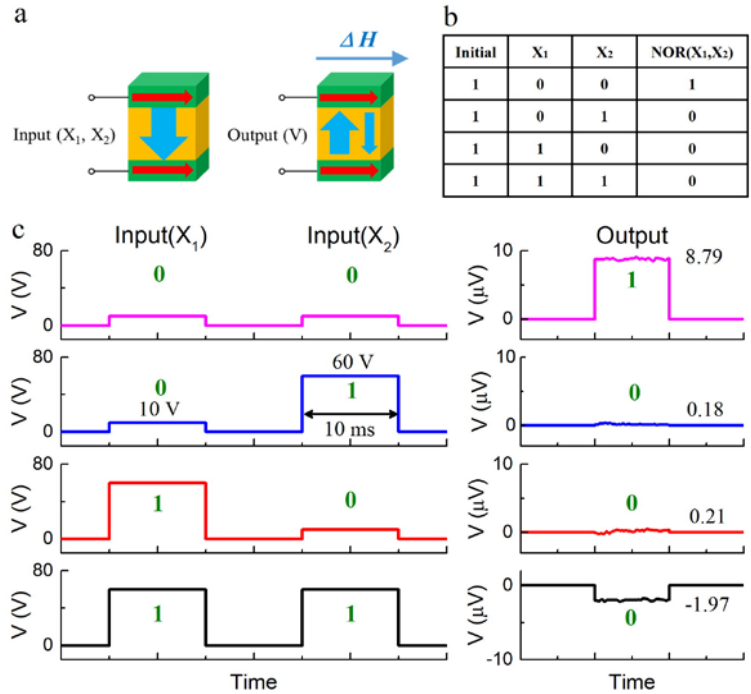

**Figure 4.** Nonvolatile NOR logic based on a single memtranstor – the second mode. (a) The schematic of the device structure and logic operations. After initialization, two sequent voltage pulses $X_1$ and $X_2$ were inputted into the device to do the computation. The low input (10 V) is set as logic "0" and the high input (60 V) as "1". The result is read out by applying a small in-plane magnetic field ($\Delta H$) to generate the output voltage via the ME effect. (b) The truth table of NOR operation. (c) Experimental results obtained on the Ni/PMN-PT/Ni memtranstor demonstrating the NOR operation. The output high ME voltage is set as logic "1" and low/negative as logic "0". Inputting "0" does not alter the direction of P so that the output ME voltage remains positive high. Inputting "1" partially reverses the direction of P so that the output ME voltage becomes low.

In addition, we demonstrate NAND logic function based on a single memtranstor. NAND is another universal Boolean logic function of two binary inputs, where the output is always "1", except for the inputs being both "1", in which case the output is "0". The principle and experimental results are shown in Fig. 5. After Initialization, two sequent voltage pulses $X_1$ and $X_2$ were inputted into the device to do the computation. The low input (10 V) is set as logic "0" and the high input (58 V) as logic "1". The output positive ME voltage is set as logic "1" and negative as "0". Inputting low voltage does not alter the direction of $P$ so that the output ME voltage remains positive high (logic "1"). Inputting a single 58 V partially reverses the direction of $P$ so that the output ME voltage reduces but still retains positive (logic "1"). Only after inputting two high voltages (58 V), the majority of $P$ is reversed so that the output ME voltage becomes negative (logic "0"). In this way, the computations fulfill the true table of the NAND logic.

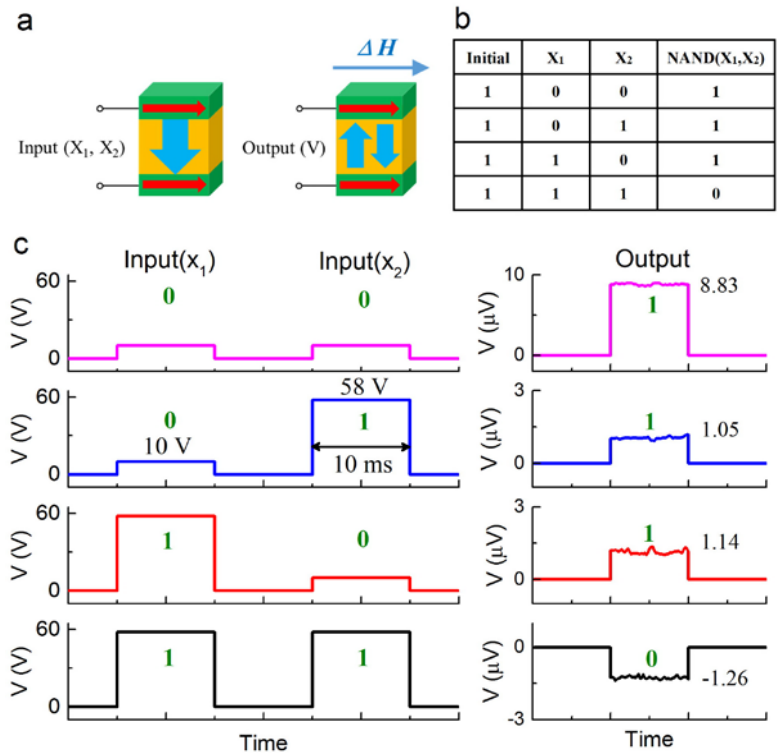

**Figure 5**. Nonvolatile NAND logic based on a single memtranstor. (a) The schematic of the device structure and logic operations. After initialization, two sequent voltage pulses $X_1$ and $X_2$ were inputted into the device to do the computation. The low input (10 V) is set as logic "0" and the high input (58 V) as logic "1". (b) The truth table of NAND operation. (c) Experimental results obtained on the Ni/PMN-PT/Ni memtranstor demonstrating the NAND operation. The output positive ME voltage is set as logic "1" and negative as logic "0". Inputting "0" does not alter the direction of $P$ so that the output ME voltage remains positive high. Inputting a single "1" partially reverses the direction of $P$ so that the output ME voltage reduces but still retains positive. Only after inputting two "1", the majority of $P$ is reversed so that the output ME voltage becomes negative.

The above experimental results successfully demonstrate that nonvolatile logic gates such as NOR and NAND can be implemented by using a single memtranstor. Compared with conventional CMOS logic gates, the logic operations in the memtranstors are quite different. First, the memtranstor-based logic gates are nonvolatile whereas the CMOS logic gates are volatile. Nonvolatile behavior could greatly suppress the static power dissipation, which is a key concern in CMOS logic due to the subthreshold leakage of transistors. The second difference is the sequential operation in the memtranstor logic, instead of the usually parallel operation in CMOS where one or more inputs are applied to a logic gate within the same clock pulse. Sequential operation is also adopted in several memristor-based logic devices [17,23,24,26,27]. Though sequential operation may take a longer time to complete a single logic operation, on the other hand, it is balanced by a significant reduction of occupied area, namely a single memtranstor as opposed to two or more CMOS transistors. Just like the memristors that could enable memory and computing integrated on the same chip to reduce interconnect delays due to the transfer of data from the memory circuit to the logic circuit, the memtranstor employing the nonlinear ME effects opens another promising way toward logic-in-memory computing system.

## IV. CONCLUSIONS

The memtranstor that correlates directly charge and magnetic flux via the nonlinear ME effects is considered as the fourth memelement in addition to memristor, memcapacitor, and meminductor. It provides another promising candidate for developing nonvolatile devices. In this work, both multi-level nonvolatile memory and nonvolatile logic functions have been successfully demonstrated based on a single memtranstor made of the Ni/PMN-PT/Ni heterostructure. The combined functions of both memory and logic could enable the memtranstor as elements for computing systems beyond von Neumann architecture. In the future, the potential of the memtranstor employing the nonlinear ME effects in generating advanced electronic devices deserves an extensive study.

**Acknowledgements**

This work was supported by the the National Key Research Program of China (Grant No. 2016YFA0300700), the National Natural Science Foundation of China (Grant Nos. 11227405, 11534015, 51371192), and the Chinese Academy of Sciences (Grants No. XDB07030200 and KJZD-EW-M05).